# Analysis of the Relationships among Longest Common Subsequences, Shortest Common Supersequences and Patterns and its application on Pattern Discovery in Biological Sequences

Kang Ning[1], Hoong Kee Ng[2] and Hon Wai Leong[2]

[1]Department of Pathology, University of Michigan, Ann Arbor, MI, USA

[2]Department of Computer Science, National University of Singapore, Singapore

kning@umich.edu, {nghoongk,leonghw}@comp.nus.edu.sg

## Abstract

*For a set of mulitple sequences, their patterns,Longest Common Subsequences (LCS) and Shortest Common Supersequences (SCS) represent different aspects of these sequences' profile, and they can all be used for biological sequence comparisons and analysis. Revealing the relationship between the patterns and LCS/SCS might provide us with a deeper view of the patterns of biological sequences, in turn leading to better understanding of them. However, There is no careful examinaton about the relationship between patterns, LCS and SCS. In this paper, we have analyzed their relation, and given some lemmas. Based on their relations, a set of algorithms called the PALS (PAtterns by Lcs and Scs) algorithms are propsoed to discover patterns in a set of biological sequences. These algorithms first generate the results for LCS and SCS of sequences by heuristic, and consequently derive patterns from these results. Experiments show that the PALS algorithms perform well (both in efficiency and in accuracy) on a variety of sequences. The PALS approach also provides us with a solution for transforming between the heuristic results of SCS and LCS.*

## 1. Introduction

DNA and protein sequences in organisms contain patterns that are strongly conserved through evolution because they are highly likely to be involved in vital biological functions. The coding regions in nucleotide sequences for DNA, for example, are highly conserved where they are important for gene expression or as marker for promoter binding sites. In proteins, conserved regions may be involved in important areas such as to define its most important fold, its many binding sites or simply its reaction to an enzyme.

The frequent occurrence of the same pattern in biological sequences usually indicates that the sequences are biologically related (i.e., they contain similar motifs, such as transcription factor biding sites in DNA sequences), and it is assumed that these important regions are better conserved in evolution. These patterns are possibly related to an important function of a set of sequences, and are important factors in sequences classification. For these reasons, pattern discovery has become one of the important problems in bioinformatics.

For the problem of deriving patterns in a set of biological sequences, it is given as input a set of (related or partially related) sequences, and the goal is to find a set of patterns that are common to all or most of the sequences in the set. A good algorithm for this problem should output patterns that are of high sensitivity and specificity. The problem of pattern discovery received wide attention in the literature [1-4].

It should be noted that besides patterns, there are two other closely related terminologies for multiple sequences, namely Longest Common Subsequence (LCS) and Shortest Common Supersequence (SCS). Given two sequences $S = s_1...s_m$ and $T = t_1...t_n$, $S$ is the subsequence of $T$ (or, $T$ is the supersequence of $S$) if for each $1 \leq j \leq m$, $1 \leq i_1 < i_2 < ... < i_m \leq n$, $s_j = t_{i_j}$. Given a set of sequences $S^+ = \{S_1, S_2, ..., S_k\}$, the LCS (SCS) of $S^+$ is the longest (shortest) possible sequence $T$ such that it is a subsequence (supersequence) of each and every sequence in $S$ at the same time.

We emphasize that given a set of sequences, their LCS, SCS and pattern are related. They represent different aspects of these sequences' profile, and they can all be used for biological sequence comparisons and analysis. However, not much research has been done to analyze their relationships and harness this relationship for pattern discovery problems in an effective manner.

In this paper, our main contribution is in the analysis of the relationships among LCS, SCS and patterns for a profiled set of input sequences. We state that: every pattern (without wildcards) is a subsequence of SCS, and all patterns (without wildcards) are common subsequences at most as long as LCS.

Based on these analysis of relationships, we subsequently propose the PALS and PALS* algorithms to derive patterns from the LCS and SCS of the input sequences. The PALS algorithm based on LCS (PALS-LCS) will first generate the approximate longest common subsequence of the input sequences by heuristic, map it back to the input sequences and derive the patterns from the results. As for the algorithm based on the shortest common supersequence (PALS-SCS), our method will also first generate the approximate SCS of the input sequences by heuristic, map the input sequences to it and derive the patterns from the results. The PALS* algorithms improve the PALS algorithms by incorporating pattern-driven approaches.

Section 2 describes some of the existing work on pattern discovery, identification of LCS and SCS in biological sequences. We then present the problem formulation and analyses in Section 3, followed by the PALS and PALS* algorithms in Section 4. In Section 5, experiment settings and results are shown to measure the performance of our algorithms. Section 6 concludes our paper with possible future work and extensions.

## 2. Existing work

There is already much research on pattern discoery from a set of sequences [1-3]. Pattern discovery algorithms can be generally divided into two categories [2]. The first is the pattern-driven (PD) approach, which is based on enumerating candidate patterns in a given solution space and picking out the ones with high fitness. The advantage of PD is that it is possible to guarantee finding the best patterns of limited size, almost regardless of the total length of the sequences. The second is the sequence-driven (SD) approach, which tries to find patterns by comparison of the given sequences and to look for similarities between them. The SD approach is able to discover patterns of almost any arbitrary size, but in general it is impossible to guarantee optimality of the results without greatly decreasing efficiency. This is because the precise comparison of multiple sequences (such as multiple sequence alignment) is NP-hard. Hence SD algorithms tend to be based on heuristics.

Some of the well-known pattern discovery algorithms include the TEIRESIAS algorithm [1] and Pratt algorithm [6], which are algorithms that combine PD and SD approaches. In the TEIRESIAS algorithm [1], all elementary (short) patterns are found in the scanning phase, and then these elementary patterns are glued with other elementary patterns at both ends (in all possible ways using depth first search) into maximal patterns in the convolution phase. The TEIRESIAS algorithm can guarantee all patterns that appear in at least a (user-defined) minimum number of sequences. The patterns used in TEIRESIAS have the format Y..A, which match any sequences containing a substring starting with Y, followed by 2 arbitrary characters, followed by A. A drawback of this algorithm is that it does not handle flexible gaps, and only allow sole residue (a single alphabet) to occupy a single position. The Pratt algorithm [6] is designed for pattern discovery in a set of protein sequences. It aims to find at least $m$ in the given $n$ sequences according to a fitness measure based on minimum description length (MDL). The patterns used in Pratt have the format Y-x(1,3)-[AC], which match any sequences containing a substring starting with Y, followed by 1 to 3 arbitrary characters, followed by either A or C. Recently, Ng and Shinohara [7] had proposed the minimal multiple generalization (MMG) method to find patterns in very scarce sequence samples. The patterns used in MMG have the format Y*A, which match any sequences containing a substring starting with Y, followed by any number of arbitrary characters (but usually of a limited length due to biological constraints), followed by A. This algorithm is proven empirically to derive patterns close to known patterns, but it requires specific initial patterns to be used.

Despite the vast effort devoted to pattern discovery in biological sequences, current algorithms still face the problem of significant degradation of performance with the increasing number of sequences [8].

For the LCS and SCS problems, it is a fact that the SCS problem is NP-hard. Owing to that, a $|\Sigma|$-approximation is produced by using the *periodic supersequence* $S_{ps} = (\alpha_1\alpha_2...\alpha_{|\Sigma|})^K$, where $\Sigma = \{\alpha_1, \alpha_2, ...\alpha_{|\Sigma|}\}$. Better results are produced by simple *Sum Height* (SH) or *Min Height* (MH) algorithms [9] which examine the characters in the sequences one by one (character-by-character approach), or variants of them that involve methods like randomization, look ahead, etc. In our recent paper [10], we have shown that none of these heuristic algorithms on short sequences are constantly better than other algorithms, and we have proposed the LAP algorithm that outperforms other algorithms in most of the cases. The LAP algorithm is a post-processing algorithm based on character-by-character approach that first generates a synthesis sequence for all of the sequences and then tries to shorten this synthesis sequence while preserving the common supersequence property. Currently, the LAP algorithm is one of the best heuristic algorithms (in terms of results length and efficiency) for the SCS problem.

The LCS problem is also NP-hard. To tackle the LCS problem, we have proposed the *Deposition and Extension* algorithm [11], in which we first generate a common subsequence for a set of sequences based on searching for common characters in a certain range of every sequences, then concatenate these common characters to form a common subsequence, and subsequently extend this common subsequence to get the result. This is also based on character-by-character approach. The Deposition and Extension algorithm is currently one of the best heuristic algorithms (also in terms of results length and efficiency) for the LCS problem.

## 3. The relationship between patterns, LCS and SCS

In this section, we first formulate the problem and describe the terminologies used. Next we examine the relationship between patterns, LCS and SCS. We observe that patterns, LCS and SCS of a set of sequences are highly related, which leads to the intuition behind the PALS algorithms.

### 3.1. Terminologies and problem formulation

A pattern, without wildcard, should be a subsequence of one or more sequences in $S$. A pattern of a set of sequences $S = \{S_1, S_2 \ldots S_n\}$ can be represented as sequence $P = p_1p_2...p_m$, in which every character $p_i$ is in alphabets set $\sum \cup \{*\}$. $\sum$ include all of the possible alphabets in the set of sequences S, and '*' matches any string of length 0 or more over $\sum$. A **language** $L(P)$ defined by pattern $P$ contains all of the sequences that can be derived from $P$ by substituting '*' with a sequence composed of characters in $\sum$ (with maximum length of $L$ and average length of $l$). A pattern $P'$ is said to be *derived* from pattern $P$ if $P \neq P'$ and $P$ is a substring (without wildcard) of $P'$. For example, *AT* and A*T* are derived from *T*. For a pattern $P'$ derived from $P$, $P'$ is said to be more **specific** than $P$. A pattern $P$ is called **maximal** on sequences set $S$ if and only if there is no pattern $P'$ that is more specific than $P$, and that $S \cap |L(P)| = |L(P')| \cap S$.

The **PD approach** for deriving the patterns from a set $S$ of sequences is formally defined as follows: Given a set $S = \{s_1, s_2, \ldots, s_n\}$ of $n$ sequences, and parameters $m$ ($m \leq n$) and $k$ (optional), find maximal patterns of length $k$ that appear in at least $m$ sequences (support). The **SD approach** for deriving the patterns from a set $S$ of sequences is formally defined as follows: Given a set $S = \{s_1, s_2, \ldots, s_n\}$ of $n$ sequences, and sets $P^+ = \{P_i \mid P_i$ is the pattern set for $S_i, 1 \leq i \leq n\}$, for any $j$ and $k$ ($j \neq k$) between 1 and $n$, combine the sets $P_j$ and $P_k$ recursively into new set $P$ such that it matches $P_j \cup P_k$ with high fitness.

For a set of sequences $S = \{s_1, s_2, \ldots, s_n\}$, and a set of patterns $PS = \{P_1, P_2, ..., P_q\}$, a sequence $s_j$ is covered by $PS$ if it is in $L(P_i)$ for at least one $P_i \in PS$. The **sensitivity** of $PS$ is defined as the number of sequences in $S$ covered by $PS$, over $|S|$. The **specificity** of $PS$ is defined as the number of sequences in $S$ covered by $PS$, over the size of languages $|\sum\{L(P_i)|P_i \in PS\}|$.

$$\text{Sensitivity} = \frac{|\sum_i \{L(P_i)\} \cap S|}{|S|} \quad (1)$$

$$\text{Specificity} = \frac{|\sum_i \{L(P_i)\} \cap S|}{|\sum_i \{L(P_i)\}|} \quad (2)$$

Note that there is limitation on the length of the language, which is application dependent. In reality, there are much fewer sequences compared to the size of languages, since not all combinations of symbols in the language have real biological meaning. Moreover, since the size of language is large, we have used the log ratio of the specificity in the paper.

$$LS = -\log_{10}(\text{Specificity}) \quad (3)$$

The closer the $LS$ value is to 0, the better the specificity of the algorithm.

### 3.2. Relationship

As aforementioned, given a set of sequences, LCS, SCS and patterns are three different aspects of a profile of these sequences, and they are all used in biological sequence comparisons and analyses. Therefore, they could be highly correlated, and there could be a transformation function among them.

We now give some lemmas about their relationships as depicted in Figure 1.

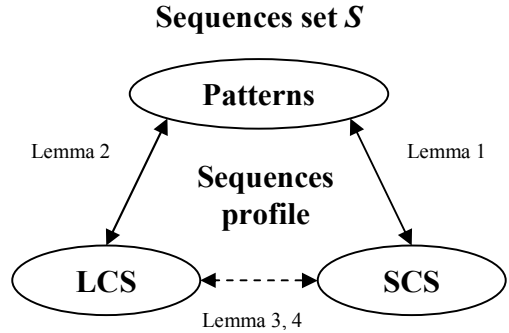

**Figure 1. Patterns, LCS and SCS are three different aspects for the profile of a set of sequences.**

**Lemma 1.** For a set $S$ of sequences, patterns without wildcard (*) are subsequences of the SCS ($S_{SCS}$) of $S$.

**Proof.** Suppose there exists a pattern $P_j$ without wild card which is not a subsequence of $S_{SCS}$. Since $P_j$ is a subsequence of some sequence $S_i \in S$, $S_i$ is also not a subsequence of $S_{SCS}$. However, by definition, $S_i$ should be a subsequence of $S_{SCS}$. This leads to a contradiction, and thus every pattern $P_j$ must be a subsequence of $S_{SCS}$. ∎

Based on the same lemma, a pattern (without wildcard) can be regarded as a subsequence of SCS which is composed of some common substrings of SCS padded with wildcards.

Therefore, given a heuristic algorithm that gives results close to exact SCS, work on deriving patterns of a set of sequences from the heuristic result of SCS is promising.

**Lemma 2.** For a set $S$ of sequences, all of the patterns without wildcard (*) are common subsequences, and they are no longer than the exact LCS ($S_{LCS}$) of $S$. Common subsequences can be seen as the patterns without wildcard. ∎

This lemma is obvious. Actually, a pattern (without wildcard) is a common subsequence. The longer the heuristic LCS results that can be generated, the more specific a pattern will be.

Based on this lemma, it is possible that patterns be derived from the heuristic result of LCS.

**Lemma 3.** Any heuristic SCS and LCS, and at least one pattern contain one or more substrings in common. ∎

This lemma can be easily derived from Lemma 1 and Lemma 2. This lemma indicates that SCS, LCS and patterns are related since they have common components. Another factor tied with their relationship is the orders of these substrings in LCS, SCS and patterns.

**Lemma 4.** If there exist heuristic SCS, LCS that contain a set of substrings $ss_1 \ldots ss_n$ in sequencial order, then there is at least one pattern that also contains these substrings in sequencial order. ∎

This lemma is apparent. It can be directly derived from Lemma 2 and Lemma 3.

From the algorithmic aspect, it is also interesting to note that the abovementioned heuristic algorithms for LCS and SCS based on character-by-character approach have inexplicitly considered patterns. In each of the search range specified in a character-by-character approach, the common character that is searched for is actually the patterns in the range. Thus, designing algorithms for pattern discovery from heuristic results of LCS and SCS follows naturally.

The heuristic algorithms for the LCS and SCS problems in the literature are generally sequence-driven (SD) algorithms. Since the precise comparison of multiple sequences (such as multiple sequence alignment) is NP-hard, it is known that the SD approach is impossible to guarantee optimality of the results without greatly decreasing efficiency.

However, current heuristic algorithms for the comparison of multiple sequences, such as computation of the LCS and SCS of a set of many sequences are becoming more efficient and accurate [11, 12]. Therefore, it is possible to devise effective pattern discovery algorithms based on these recent developments. The PALS algorithms that we have proposed adopted both the SD and PD approaches. They comprise two algorithms: the first one, PALS-LCS, is based on the heuristic result of LCS of the sequences, and the second one, PALS-SCS, is based on the heuristic result of SCS of the sequences.

## 4. The PALS algorithms

Based on the analysis of the relations of LCS, SCS and patterns, we have proposed a set of algorithms, the PALS (PAtterns by Lcs and Scs) algorithms and PALS* algorithms. The PALS algorithms proposed by us are based on the SD approach; while the PALS* algorithms improve upon the PALS algorithms by incorporating PD approaches.

### 4.1. PALS-LCS: Algorithm based on LCS

Using the relationship between patterns and LCS, we propose an algorithm to find patterns based on heuristic result of LCS of the given sequences.

Given a set $S$ of sequences, the heuristic result of the LCS of the sequences, denoted as $LCS_A(S)$, is the longest common subsequence of all the sequences in set $S$ given by heuristic algorithm $A$. In this study, we have adopted the post-processing algorithm of [11] (*Deposition and Extension* algorithm) to derive $LCS_A(S)$. The process for this post-processing is to first generate a common subsequence (template) for a set of sequences by the deposition method similar to those used in oligos synthesis, and then extend this common subsequence to get the heuristic result of LCS. We refer readers to [11] for details. In the following part, the heuristic SCS algorithm will refer to this algorithm.

After $LCS_{DepExtn}(S)$ is derived, we map (align) this $LCS_{DepExtn}(S)$ back onto every sequence in the set $S$ to derive different patterns. For example, suppose $S$ = {ACGT, CGGT, CGTC} and $LCS_{DepExtn}(S)$ = CGT, then mapping CGT to AGCT result in *CG*T. Thus the patterns are {*CG*T, CG*T, CG*T*}. We then generate the longest common *substring* of these patterns as the final pattern. Continuing the example, the pattern of sequences in $S$ is the longest common substring of {*CG*T, CG*T, CG*T*}, which is CG*T. Finally, a '*' is prepended and appended to this result to obtain the final pattern *CG*T** (refer to Section 5.1 for more details). Figure 3 lists the algorithm and Figure 3 illustrates the example.

```
PALS-LCS(S)
// input: sequences set S
// output: R, the pattern for all sequences
//         in S
begin
  L ← LCS_DepExtn(S);
  P ← Patternize_α(S,L);
  R ← LCSubstring(P);
  return R;
end;

Patternize_α(L)
// input: LCS for sequences set, L
//        sequences set S
// output: generated patterns from L
Begin
  for i ← 1 to |S|
    P ← {P, map L to S_i};
  return P;
end;

LCSubstring(P)
// input: patterns set P
// output: the longest common substring of
//         patterns in P
begin
  s ← find longest common substring of P;

  // ensure first and last char is *
  if s[1] != '*' then s ← '*' ∪ s;
  if s[length(s)] != '*' then s ← s ∪ '*';

  return s;
end;
```

**Figure 3. PALS-LCS: An algorithm to find patterns based on LCS.**

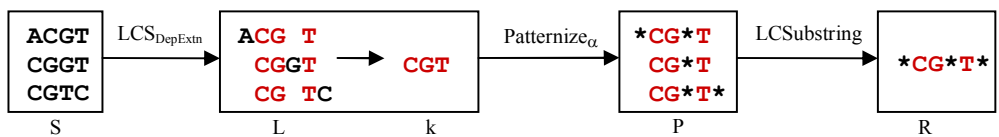

**Figure 2. An example of the PALS-LCS algorithm.**

Note that since there maybe multiple substrings for a set of patterns, the number of patterns by PALS-LCS can be one or more.

Suppose the inputs are $n$ sequences with the length of at most $k$, and the alphabet set $\sum$. The time for obtaining LCS by heuristics is $O(kn|\sum|)$, for devising patterns is $O(kn)$, and for substrings generation is $O(kn)$. Therefore, the total time complexity of PALS-LCS is $O(kn|\sum|)$. The space complexity is $O(kn|\sum|)$.

## 4.2. PALS*-LCS: Algorithm based on LCS with fewer wildcards

How to generate patterns with as few wildcards as possible is a very important issue for high quality pattern discovery, since fewer wildcards translates to reduced size of languages and more specific pattern.

The PALS*-LCS algorithm is the improved PALS-LCS algorithm with two post-processing steps:

(a) firstly, PALS*-LCS tries to remove redundant wildcards in patterns while keeping the patterns property intact. Given any pattern generated by PALS-LCS, PALS*-LCS performs the following post-processing sub-steps:

(i) if by removing a wildcard in this pattern, the remaining pattern is still the pattern for the sequences, then this wildcard is removed (note that this removal happens with a small chance);

(ii) if by reversing the neighboring wildcard and alphabet, the number of wildcards can be reduced (for example (*$a_i$*) to ($a_i$**) to ($a_i$*)), then these wildcards are reduced.

(b) secondly, PALS*-LCS applies a PD (pattern driving) approach to iteratively reduce the wildcards and increase the number of alphabets in the patterns, while still ensuring that these patterns appear in a minimum number of sequences. Then the pattern with the best specificity is selected.

We use an example with $S$ = {ACGT, CGGT, CTGC}, the PALS-LCS algorithm returns the pattern *C*G*. If we allow one of the sequences to mismatch with the pattern, then this pattern can be improved to *CGT*, which has higher specificity.

It is easy to see that the time complexity of the PALS*-LCS algorithm is $O(k^2n|\sum|)$, and the space complexity is $O(kn|\sum|)$.

## 4.3. PALS-SCS: Algorithm based on SCS

The algorithm based on the heuristic result of the SCS of the sequences is similar to that based on LCS.

Given a set $S$ of sequences, the heuristic result of the SCS of the sequences, denoted as $SCS_A(S)$, is the shortest common supersequence of all the sequences in set $S$ given by heuristic algorithm $A$. In this study, we have adopted the post-processing algorithm of [12] (*Deposition and Reduction* algorithm) to derive $SCS_A(S)$. This algorithm first generates a template pool – a small set of SCS templates (or templates, in short). Each template is a common supersequence of the SCS instance $S$. The reduction process shortens these templates by attempting to remove some characters while preserving the common supersequence property. We refer readers to [12] for details. In the following part, the heuristic SCS algorithm will refer to this algorithm.

After $SCS_{DepRedn}(S)$ is derived, we map every sequence in the set $S$ onto this $SCS_{DepRedn}(S)$ to derive different patterns. For example, suppose $S$ = {ACGT, CGGT, CGTC}, and $SCS_{DepRedn}(S)$ = ACGGTC, then the patterns are {*ACG*T**, **CGGT**, **CG*T*}. We then generate the LCS of these patterns as the final pattern by using the heuristic algorithm for the LCS problem. For the same example, the pattern of sequences in $S$ is the LCS of {*ACG*T**, **CGGT**, **CG*T*}, which is *CG*T*. Figure 4 lists the algorithm and Figure 5 illustrates the example.

Note that by using the heuristic algorithm for the LCS problem to get the final results, the patterns that we have obtained are maximal patterns.

Suppose the inputs are $n$ sequences with the length of at most $k$, and the alphabet set $\sum$. The time complexity to post-process SCS is $O(kn|\sum| + k^2n)$, to devise patterns is $O(kn)$, and to generate subsequences is $O(kn|\sum|)$. Therefore, the total time complexity of PALS-SCS is $O(kn|\sum| + k^2n)$. The space complexity of the algorithm is $O(kn|\sum|)$.

```
PALS-SCS(S)
// input: sequences set S
// output: R, the pattern for all sequences
//         in S
begin
  L ← SCS_DepRedn(S);
  P ← Patternize_β(L);
  R ← LCS_DepExtn(P);
  return R;
end;

Patternize_β(L)
// input: SCS set L
// output: generated patterns of L
begin
  k ← find SCS of L;
  for i ← 1 to |L|
    P ← P ∪ {map L_i to k};
  return P;
end;
```

**Figure 4. PALS-SCS: An algorithm to find patterns based on SCS.**

**Figure 5. An example of the PALS-SCS algorithm.**

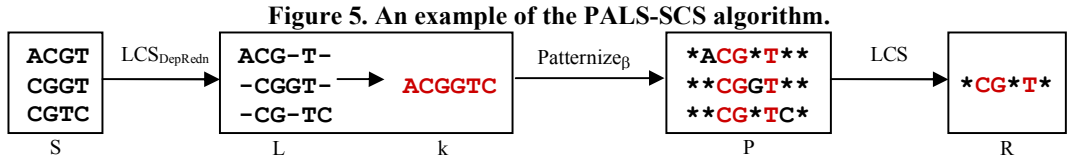

### 4.4. PALS*-SCS: Algorithm based on SCS with fewer wildcards

The PALS*-SCS algorithm is the improved PALS-SCS algorithm with post-processing. The post-processing is the same as in PALS*-LCS. However, note that in step (a)(i), the chance of wildcard removal is much higher than that for PALS*-LCS. This is easy to see just by comparing the pattern generation process between PALS-SCS and PALS-LCS. Again, the time complexity of the PALS*-SCS algorithm is $O(k^2n|\Sigma|)$, and space complexity is $O(kn|\Sigma|)$.

It is easy to see that the sensitivity of both PALS-LCS and PALS-SCS is 100%, since the patterns generated cover all of the sequences in the dataset. On the other hand, by integrating more PD approaches into the algorithms PALS*-LCS and PALS*-SCS, we can achieve higher specificity at the cost of lower sensitivity.

### 4.5. Transformation between LCS and SCS though patterns

The PALS approach also provides us with a solution for transforming between the heuristic results of SCS and LCS: the heuristic results of SCS and LCS can be transformed though patterns.

The transformation is straightforward: to transform heuristic result of SCS to that of LCS, the pattern is generated based on heuristics SCS algorithm. Then the sequences in the dataset are aligned, under the guidance of patterns. This alignment results in the heuristic LCS. Theoretically, the transformation from heuristic results of LCS to SCS is a little bit more complicated, since the LCS only contains the common characters for a set of sequences, while SCS contains every character for each of these sequences (and in the same order as in these sequences). However, the transformation from heuristic results of LCS to SCS is similar to that from heuristic SCS to that of LCS in practice: After pattern generation based on heuristics LCS algorithm, all of the sequences in the dataset can also be aligned under the guidance of patterns, and this alignment results in the heuristic SCS.

### 4.6. Refinement of the heuristic results of SCS, LCS and patterns

Based on approach proposed in the previous section, we think that the iterative refinement of heuristic SCS, LCS and patterns may result in better SCS, LCS and patterns. The iterative refinement is performed as this: For heuristic results of LCS and SCS, we iteratively transform one to another, if either (a) the length of LCS or SCS is improved, or (b) specificity or sensitivity is improved in the whole transformation process. This process is terminated when there is no improvement in the process.

Take a slightly different example with $S$ = {ACGT, CGGT, CTGC}. The heuristic SCS, $SCS_{DepRedn}(S)$ = ACTGGTC, and the pattern is *C*G*, which result in a heuristic LCS of CG. On the other hand, the heuristic LCS directly generated from sequences is G. Therefore, the heuristic LCS from heuristic SCS is better.

Another refinement of the results is to generate more than one heuristic result of LCS and SCS. This is beneficial for generation of good patterns, and the heuristic results of LCS and SCS can also benefit from transformation. For example, in the above example, if multiple heuristic LCS are generated (CT, CG, G), then CG is already one of them. This makes the patterns from LCS, *C*T* or *C*G*, better patterns than *G*.

## 5. Experiments

In experiments, we have focused on using PALS and PALS* algorithms for pattern discovery. Scrutinizing the PALS and PALS* algorithms, it is apparent that two questions arise and need to be answered: firstly, how many of the sequences in the set are covered by the patterns derived; and secondly, what proportion of the known patterns is covered by the patterns derived. To answer these questions on sensitivity and specificity, we performed experiments on simulated and real sequences. We have also analyzed the refinement of heuristic results.

### 5.1. Experiment settings and datasets

To prove the effectiveness of our algorithms on general datasets, we have obtained simulated and real DNA sequences for experiments. Analysis on protein sequences was also performed, with similar results. The simulated DNA sequences contain random sequences generated by our in-house random DNA generator. For each value of the number of sequences N = {10, 100, 1000}, and sequences length K = {100, 1000}, we generated 10 random datasets of DNA sequences. Then each algorithm is run on these 10 instances to get the average results. The real datasets used include DNA sequences obtained from the DBTBS database [13] with known consensus patterns. Another dataset that we used is a subset of protein sequences retrieved from the IPI Human database [14].

Our algorithm is implemented in C++ and Perl. To derive a list of the longest common substrings for given patterns, we used the Perl Tree::Suffix module. The experiments were performed on a Linux machine with 3.0GHz CPU and 1GB RAM. For pattern discovery, the PALS algorithms are compared with the TEIRESIAS [1], Pratt [6] and MMG [7] algorithms. Default values were used for all the other algorithms.

Since each of these algorithms has a different definition of wildcards, it is difficult to calculate the exact size of language, Here we assume that for a specific dataset, the corresponding language have a same average length, regardless of the algortihm used. Specificly, for PALS algorithms, as well as the MMG algorithm, it is assumed that the average length of sequences being analyzed is $l$, average length of patterns without wildcard is $p$, and average number of wildcards is $q$. Then for every pattern, each wildcard '*' found between two alphabets in the pattern can represent substring of length $(l-p)/q$, and the language size is $|\Sigma|^{(l-p)}$. For example, when *AC*T* and *AC* represent the same number of DNA sequences of average length of 7, then the language represented by *AC*T* is $4^{(7-3)}$=64, and the language represented by *AC* is $4^{(7-2)}$=128. The former is more specific than the latter on this set of sequences. Since the average length of sequences is known in advance, this calculation can accurately estimate the size of languages. We also assume (for TEIRESIAS) that each wildcard '.' represents $|\Sigma|$ characters. The Pratt algorithm has similar scheme so that the language size can also be easily computed by multiplication arithmetic. As we have previously mentioned, *LS* is used instead of Specificity.

**Table 1. Analysis of specificity and sensitivity of the patterns, as well as running time by PALS and PALS* algorithms. "LS" represent log(specificity), and "No. of Pattern" represent the average number of patterns from the repective algortihms.**

| Base Methods | No. of Seqs | Length of Seqs | No. of Pattern | No. of Covered Seqs | PALS | | | PALS* | | |
| --- | --- | --- | --- | --- | --- | --- | --- | --- | --- | --- |
| | | | | | LS | Sensitivity (%) | Time (secs) | LS | Sensitivity (%) | Time (secs) |
| LCS | 10 | 100 | 2.7 | 10 | 2.30 | 100 | 0.8 | 2.30 | 100 | 12 |
| | 10 | 1000 | 4.0 | 10 | 3.32 | 100 | 80.5 | 3.30 | 100 | 132 |
| | 100 | 100 | 1.6 | 100 | 3.65 | 100 | 4.4 | 3.65 | 100 | 15 |
| | 100 | 1000 | 3.9 | 100 | 4.66 | 100 | 548.3 | 4.65 | 100 | 957 |
| | 1000 | 100 | 2.6 | 1000 | 7.44 | 100 | 46.0 | 7.41 | 100 | 81 |
| | 1000 | 1000 | 4.1 | 1000 | 7.84 | 100 | 6040.1 | 7.80 | 100 | 14134 |
| SCS | 10 | 100 | 1 | 10 | 2.38 | 100 | 529 | 2.30 | 100 | 800 |
| | 10 | 1000 | 1 | 10 | 3.07 | 100 | 7910 | 3.30 | 100 | 12000 |
| | 100 | 100 | 1 | 100 | 3.81 | 100 | 941 | 3.65 | 100 | 2280 |
| | 100 | 1000 | 1 | 100 | 4.64 | 100 | 12481 | 4.65 | 100 | 22200 |
| | 1000 | 100 | 1 | 1000 | 5.50 | 100 | 1239 | 5.44 | 100 | 2160 |
| | 1000 | 1000 | 1 | 1000 | 5.81 | 100 | 15866 | 5.40 | 100 | 34500 |

### 5.2. Results of PALS and PALS* algorithms

We had first analyzed the specificity and sensitivity of the patterns derived by PALS algorithms on simulated sequences. For PALS-LCS, since we have obtained a set of longest common substrings, the number of patterns is always larger than 1 whereas for PALS-SCS, there is only one heuristic pattern result. The total running time is also examined. For simplicity, we only show the results of PALS* with sensitivity=1. More analysis on different sensitivities for PALS* algorithms will be shown later.

Results in

Table **1** show that the sensitivities of the patterns derived by PALS algorithms are high. The sensitivities of both algorithms achieve 100% for different datasets. However, the specificities are low. This is probably due to the definition of the size of the languages (which includes much more sequences than is in the biological sense). The specificity of PALS-SCS is comparable to that of PALS-LCS for small dataset (N ≤ 100), but is higher than PALS-LCS for large datasets (N ≥ 1000). This is because the SCS of a set of sequences contains more common information about sequences set than the LCS of those sequences.

For the running time of PALS-LCS, we can see from the results that even for large sequence sets with many (1000) long (100) sequences, the processing time is less than 1 minute. For larger datasets with more than 100 sequences with length 1000, the time needed is a few minutes. The running time of PALS-SCS is much slower than PALS-LCS for the same datasets. This is due to the fact that the time complexity of PALS-SCS is greater than PALS-LCS, especially for long sequences.

Generally, PALS* algorithms perform better than PALS algorithms (

Table **1**). This is likely due to the post-processing in PALS* algorithms. More specifically, PALS*-LCS improve very little from PALS-LCS, while the difference between PALS*-SCS and PALS-SCS is large. Because of post-processing, the PALS* algorithms are slower.

Next, we generated patterns for several sets of real sequences with PALS and PALS*, and compared them with their respective known consensus patterns. Again, we only show the results of PALS* with sensitivity=1. Results in

Table **3** and Table 4 show that the patterns by PALS-LCS and the patterns by PALS-SCS have certain similarity with the known consensus patterns for the same dataset. For example, the known consensus pattern for SigD is *TAAA*GCCGATAT*, and the pattern by PALS-SCS is *TAAA*T*T*CA*A*A*AA*. A significant fragment, "TAAA" is found by PALS-SCS. For the patterns generated by PALS-SCS, we also observed that many of them are supersequences of the corresponding known consensus patterns. For the patterns generated by PALS-LCS, there is 100% sensitivity for every dataset, and the specificity values are not very low.

Most of the results of PALS* algorithms are the same as those of PALS on these real sequences. On SigD dataset, the pattern by PALS*-SCS is *TAAA*A*A*AA*A*A*A*AA*. Since it is assumed that the length of the language that the patterns represent is a fixed value, this pattern is more specific than the result of PALS-SCS.

Further, we analyzed the results of the PALS* algorithms with larger specificity and smaller sensitivity. The results are shown in Table 2. It is obvious that by reducing the sensitivity of the results, both PALS*-LCS and PALS*-SCS algorithms can effectively increase the specificity of the results, with the cost of slight decrease of sensitivity. On the other hand, in order not to increase the computational time greatly, the specificities of heuristic results are also not drastically improved.

**Table 2. Analysis of specificity and sensitivity of the patterns by PALS* algorithms with different sensitivities.**

| Base Methods | No. of Seqs | Length of Seqs | MIN(sensitivity) | | | | | |
|---|---|---|---|---|---|---|---|---|
| | | | 1 | | 0.9 | | 0.8 | |
| | | | *LS* | Sen (%) | *LS* | Sen (%) | *LS* | Sen (%) |
| **LCS** | 10 | 100 | 2.30 | 100 | 2.20 | 90 | 2.20 | 80 |
| | 10 | 1000 | 3.30 | 100 | 3.10 | 90 | 3.00 | 80 |
| | 100 | 100 | 3.65 | 100 | 3.60 | 91 | 3.60 | 80 |
| | 100 | 1000 | 4.65 | 100 | 4.55 | 90.5 | 4.45 | 81.5 |
| | 1000 | 100 | 7.41 | 100 | 7.35 | 92 | 7.23 | 81 |
| | 1000 | 1000 | 7.80 | 100 | 7.66 | 91 | 7.36 | 81.5 |
| **SCS** | 10 | 100 | 2.30 | 100 | 2.30 | 90 | 2.30 | 80 |
| | 10 | 1000 | 3.30 | 100 | 3.30 | 90 | 3.10 | 80 |
| | 100 | 100 | 3.65 | 100 | 3.60 | 91 | 3.60 | 81 |
| | 100 | 1000 | 4.65 | 100 | 4.50 | 92 | 4.40 | 81 |

| | 1000 | 100 | 5.44 | 100 | 5.31 | 90 | 5.25 | 80.5 |
|---|---|---|---|---|---|---|---|---|
| | 1000 | 1000 | 5.40 | 100 | 5.33 | 90 | 5.28 | 80 |

## 5.3. Comparison with other algorithms

Next, we compared PALS and PALS* algorithms to well-known algorithms: MMG [7], TEIRESIAS [1] and Pratt [6]. As the "gold standard", we used the results from the "regular pattern of fixed form with range specifiers" [7], which have the best reported performance. Since PALS-LCS and PALS*-LCS perform better than PALS-SCS and PALS*-SCS respectively, we only used PALS-LCS and PALS*-LCS for comparison. We show the results of PALS*-LCS with sensitivity=1 for simplicity.

Results in Table 5 show that PALS algorithms have comparable sensitivity with the TEIRESIAS, and better sensitivity than the MMG algorithm. PALS-LCS always outputs results with 100% sensitivity; these are the same for the TEIRESIAS algorithm, but the MMG algorithm output results with less sensitivity. Though PALS-LCS has low specificity, and are less accurate than the MMG algorithm, they are more accurate than the TEIRESIAS algorithm. This is because TEIRESIAS algorithm produces all the patterns that appear in at least a minimum number of sequences, so the language size is very large. PALS*-LCS algorithm's output is very close to that of PALS-LCS's. The Pratt algorithm perform better than MMG and TEIRESIAS algorithms both in specificity and sensitivity. The Pratt algorithm also performs better than the PALS*-LCS algorithm.

Apart from comparing the relative accuracies of these algorithms, we have also calculated the sensitivity and specificity of the randomly generated pseudo patterns and compared them with these algorithms. Only if such pseudo patterns have significantly lower specificity and sensitivity can we say that the pattern discovery algorithms are discriminative. The results (details not shown here) show that the pseudo patterns are far smaller than these algorithms compared, indicating these algorithms are discriminative.

We have also tried to analyze the common patterns found by these algorithms. We analyzed the results of PALS* and the Pratt algorithm (which has better specificity and sensitivity than other algorithms) on simulated sequences. It is interesting to note that though their patterns are similar at one part or another, hardly any of their results are identical. This indicates the non-optimality of current patterns discovered by these algorithms. We think that a meta-algorithm that can combine the results of these algorithms might give even better patterns.

## 5.4. Refinement of the heuristic results

The performance of refinement approach for the heuristic results of SCS, LCS and patterns are investigated next. Generally, refinement can further improve the sensitivity and specificity of the patterns generated, as well as gives longer heuristic LCS and shorter SCS on some datasets. On average the improvement is small (by 1 or 2 characters). Although achieving such improvement is important, the time used is much more than the PALS algorithms. This indicates that the current heuristic algorithms for LCS and SCS that we used have reached a limit based on the character-by-character approach, and further refinements based purely on these results are not effective.

When we used more than one heuristic result for LCS and SCS generation, it was discovered that the qualities of the resulting LCS, SCS and patterns are all improving (details not shown here). This is as expected, since more information has been gathered from these sequences. Generally, by this means, the length of LCS and SCS can be further improved by 1 to 2 characters, and the patterns can be more specific.

Table 3. Comparison between known consensus patterns and the results of PALS and PALS*

| | | PALS-LCS | | | PALS*-LCS | | |
|---|---|---|---|---|---|---|---|
| Datasets | Known Consensus Patterns | No. Patterns by LCS | LS | Sensitivity (%) | No. Patterns by LCS | LS | Sensitivity (%) |
| SigB | *AGGTTT*GGGTAT* | 3 | 10.52 | 100 | 3 | 10.52 | 100 |
| SigD | *TAAA*GCCGATAT* | 1 | 10.06 | 100 | 1 | 9.92 | 100 |
| SigE | *CATAT*CATACA*, *ATATT*CATACA* | 2 | 10.69 | 100 | 2 | 10.16 | 100 |
| SigF | *G*TA*, *GG*A*A*TA* | 2 | 9.40 | 100 | 2 | 9.34 | 100 |
| SigG | *GHATR*, *GG*CATXHTA* | 1 | 9.14 | 100 | 1 | 9.14 | 100 |
| SigH | *AGGTATT*GAATT* | 1 | 10.18 | 100 | 1 | 10.16 | 100 |
| SigL | *TGGCA*TTGCA* | 2 | 9.07 | 100 | 2 | 9.07 | 100 |
| SigW | *TGAACN*CGTA* | 2 | 10.19 | 100 | 2 | 10.19 | 100 |

Table 4. Comparison between known consensus patterns and the results of PALS and PALS*.

| Datasets | Known Consensus Patterns | Pattern by PALS-SCS | Pattern by PALS*-SCS |
|---|---|---|---|
| SigB | *AGGTTT*GGGTAT* | *G*A*A*GG*A*A*A* | *G*A*A*GG*A*A*A* |
| SigD | *TAAA*GCCGATAT* | *TAAA*T*T*CA*A*A*AA* | *TAAA*A*A*AA*A*A*A*AA* |
| SigE | *CATAT*CATACA*, *ATATT*CATACA* | *T*T*T*T*TA*A*A*A* | *T*T*T*T*TA*A*A*A* |
| SigF | *G*TA*, *GG*A*A*TA* | *G*T*TA*TT*T*A*AA*A*TA*A* | *G*T*TA*TT*T*A*AA*A*TA*A* |
| SigG | *GHATR*, *GG*CATXHTA* | *G*A*AA*AA*A*AA*T*T* | *G*A*AA*AA*A*AA*T*T*s |
| SigH | *AGGTATT*GAATT* | *A*A*A*AG*AAT* | *A*A*A*AG*AAT* |
| SigL | *TGGCA*TTGCA* | *CA*A*AC*TGGCA*C*TTGCA*T*T*AA*AG*G*GA*A*A* | CAAACTGGCACTTGCATTAAAGGGAAA |
| SigW | *TGAACN*CGTA* | *AA*AA*C*T*TT*T*TA* | *AA*AA*C*T*TT*T*TA* |

Again, this indicates that LCS, SCS and patterns are highly related. By using multiple (albeit slightly different) LCSs and SCSs, which represent the same sequences profile, the orders of characters in patterns are slightly changed, thus making the patterns more specific.

### 5.5. Efficiency

For the running time of these algorithms, we observe that even for datasets with 10 sequences each of length 100, MMG took more than 15 minutes to process, and TEIRESIAS took about 1 minute. In fact, MMG terminates after 900 secs, and there are no results for datasets that takes longer than that to process. The Pratt algorithm also takes more than 1 minute on these datasets. By comparison, PALS is much efficient than these two algorithms (

Table 1). However, for very large dataset (N > 1000, K > 1000), all of these algorithms (including PALS) take more than an hour to process except Pratt algorithm, which only needs 10 minutes.

The programs for PALS and PALS* algorithms are available upon request.

## 6. Conclusions and future work

In this paper, we have focused on the relationships of LCS, SCS and patterns for biological sequences. The investigation of this problem is very important in bioinformatics, since the patterns in biological sequences usually indicate structural or functional relationship among sequences. The contributions of this paper include (a) the observation that for a set of sequences, the sequences profile is the center around which LCS, SCS and patterns are highly related; and (b) novel algorithms to derive patterns that are based on the observation of relationship among LCS, SCS and patterns for the given sequences.

The algorithms proposed, PALS and PALS*, have high sensitivity and specificity, and they are effective in deriving patterns close to the known consensus patterns for real sequences. For PALS-SCS, the patterns generated are maximal patterns. The PALS* algorithms incorporate a post process that further improve the specificity. The sensitivities and specificities of PALS and PALS* are comparable to or higher than existing algorithms such as TEIRESIAS, MMG and Pratt on different sequences set. The PALS and PALS* algorithms are also quite efficient,

Table 5. Comparison of different algorithms on patterns of a set of sequences. A '–' indicates that the algorithm took too long to produce results.

| Datasets | | | PALS-LCS | | PALS*-LCS | | MMG | | TEIRESIAS | | Pratt | |
|---|---|---|---|---|---|---|---|---|---|---|---|---|
| Simulated sequences | Length of Seqs | No. of Seqs | Sen (%) | LS | Sen (%) | Sen (%) | Sen (%) | LS | Sen (%) | LS | Sen (%) | LS |
| Dataset 1 | 100 | 10 | 100 | 2.30 | 100 | 100 | – | – | 100 | 15.95 | 100 | 11.04 |
| Dataset 2 | 1000 | 10 | 100 | 3.32 | 100 | 100 | – | – | 100 | – | 100 | 17.06 |
| Dataset 3 | 100 | 100 | 100 | 3.65 | 100 | 100 | – | – | 100 | 18.55 | 100 | 8.24 |
| Dataset 4 | 1000 | 100 | 100 | 4.66 | 100 | 100 | – | – | 100 | – | 100 | 16.06 |
| Real sequences | Known Consensus Patterns | No. of Seqs | | | | | | | | | | |
| SigD | *TAAA*GCCGATAT* | 33 | 100 | 10.06 | 100 | 100 | 97.0 | 2.80 | 100 | 12.81 | 100 | 5.46 |
| SigE | *CATAT*CATACA*, *ATATT*CATACA* | 62 | 100 | 10.69 | 100 | 100 | 96.8 | 2.53 | 100 | 16.86 | 100 | 3.38 |
| SigH | *AGGTATT*GAATT* | 48 | 100 | 10.18 | 100 | 100 | 91.7 | 2.64 | 100 | 14.05 | 100 | 2.58 |
| SigW | *TGAACN*CGTA* | 32 | 100 | 10.19 | 100 | 100 | 96.9 | 2.82 | 100 | 12.65 | 100 | 12.69 |

and have small space complexities.

Although we had investigated the relationships among LCS, SCS and patterns, we think that a deeper understanding of their relationships is needed. One possible method is the generation of a hidden markov model (HMM) for the alignment of the sequences, and analyzing the relationship between this HMM and LCS, SCS and patterns. We think that there may be a HMM available such that there are only match and insertion states. By using such HMM, the SCS can be the concatenation of all emitted symbols, the LCS can be the concatenation of all symbols emitted by match state, and the pattern can be the LCS thus generated with wildcards in it.

Another direction of research is to improve the current algorithm so that the PALS algorithms can output more useful information about the patterns of the sequences. We are currently working on this issue.

Combining the patterns that we have discovered with biological domain knowledge such as the functions of certain sequences family, it is possible that these algorithms be applied on more bioinformatics problems such as to classify sequences, predict their functions, and find new motifs. We will also works on these interesting problems in the near future.

## Acknowledgements

We thank Dr. Yen Kaow Ng from National University of Singapore for discussion of the problem. We also thank anonymous reviewers for their insightful comments.